\documentclass[aps,prl,twocolumn,letterpaper,showpacs,nofootinbib,amsmath,amssymb]{revtex4-1}

\usepackage{bm}
\usepackage{epsfig}
\usepackage{graphicx}
\usepackage[]{hyperref}
  \hypersetup{
  unicode=false,          
  pdftoolbar=true,        
  pdfmenubar=true,        
  pdffitwindow=true,     
  pdfstartview={FitH},    
  pdfsubject={Getting the best out of T2K and NOvA},   
  pdfnewwindow=true,      
  pdfcreator={RevTeX},
  colorlinks=true,       
  linkcolor=red,          
  citecolor=blue,        
  urlcolor=blue,           
  }
\usepackage{hypcap}

\usepackage{amsmath}
\usepackage{amssymb}
\newcommand{\beq}{\begin{equation}}
\newcommand{\eeq}{\end{equation}}
\newcommand{\beqarray}{\begin{eqnarray}}
\newcommand{\eeqarray}{\end{eqnarray}}
\newcommand{\dcp}{\delta_{CP}}
\newcommand{\numu}{\nu_\mu}
\newcommand{\nue}{\nu_e}
\newcommand{\nutau}{\nu_\tau}
\newcommand{\piS}{$\pi$S}
\newcommand{\muDS}{$\mu$DS}
\newcommand{\nS}{$n$S}

\begin{document}
\title{Probing CP violation with the three years ultra-high energy neutrinos from IceCube} 

\author{Animesh Chatterjee$^1$}\email{animesh@hri.res.in}
\author{Moon Moon Devi$^2$}\email{moonmoon4u@tifr.res.in}
\author{Monojit Ghosh$^3$}\email{monojit@prl.res.in}
\author{Reetanjali Moharana$^1$}\email{reetanjalimoharana@hri.res.in}
\author{Sushant K. Raut$^3$}\email{sushant@prl.res.in}
\affiliation{\it $^1$Harish-Chandra Research Institute, Jhunsi, Allahabad 211 019 , India.}
\affiliation{\it $^2$Tata Institute of Fundamental Research, Mumbai 400 005, India.}
\affiliation{\it$^3$ Physical Research Laboratory, Navrangpura, Ahmedabad 380 009, India.}

\begin{abstract}
  The IceCube collaboration has recently announced the discovery of ultra-high 
  energy neutrino events. These neutrinos can be used to probe their production 
  source, as well as leptonic mixing parameters. 
  In this work, we have used the first IceCube data to constrain the leptonic 
  CP violating phase $\dcp$. For this, we have analyzed the data in the form of flux ratios. 
  We find that the fit to $\dcp$ depends on the assumptions made on the production 
  mechanism of these astrophyscial neutrinos. 
  Consequently, we also use this data to impose constraints on the sources of the neutrinos. 
\end{abstract}


\date{\today}
\maketitle
{\bf \underline{Introduction:}}
The study of cosmic particles, and through them the study of astrophysical 
phenomena has gradually moved up the energy scale over the last few decades.
GeV scale cosmic rays have already been observed in atmospheric 
neutrino experiments for many years. 
With the recent detection of ultra-high energy (UHE) neutrinos at 
the IceCube detector\cite{Aartsen:2013jdh,Aartsen:2014gkd}, 
we have emphatically entered the TeV regime. 
In fact, the PeV energy events seen by IceCube underscore 
the prospects of neutrino astrophysics with large telescopes. 

The first data set announced by the IceCube collaboration consists of 28 events above 25 TeV, 
detected over a period of 662 days of live time (May 2011 -- May 2012 with 79 strings, 
and May 2012 -- May 2013 with 86 strings). 
7 out of these 28 events are tracks signifying 
($\numu+\bar{\nu}_{\mu}$) charged-current (CC) events; while the other 21 are 
showers indicating 
either ($\nue+\bar{\nu}_{e}$) or ($\nutau+\bar{\nu}_{\tau}$), or ($\numu+\bar{\nu}_{\mu}$) 
neutral-current (NC) events \cite{Aartsen:2013bka}. 
This $4\sigma$ detection marked the first discovery of UHE neutrinos. Further data was collected for next one year. For the full 988 days IceCube 
collected 37 events, adding 1 track, 7 shower events and 1 was produced by a coincident pair of background muons from 
unrelated air showers that cannot be reconstructed with
a single direction and energy, to 
the previously detected 28 events including the highest neutrino of energy 2000 TeV, ever detected \cite{Aartsen:2014gkd}. 

UHE cosmic rays secondaries like photons and neutrinos 
carry information about their production (source) and propagation. 
UHE neutrinos can be produced by a wide array of astrophysical and 
cosmological processes. It may be possible to probe the mechanism of their 
production by observing them at neutrino telescopes. Moreover, as these neutrinos 
travel from their source to the earth, they oscillate. Therefore, one can 
use this information to constrain mixing in the leptonic sector. 
In this paper, we have used the first data set from IceCube to address 
questions about astrophysical neutrino production and neutrino oscillations.

Data from various neutrino oscillation experiments have constrained the 
mixing angles and mass-squared differences at the $\leq 10\%$ 
level \cite{Capozzi:2013csa}. 
However, the value of the CP violating phase $\dcp$ is not constrained by the 
data. The measurement of $\dcp$ is one of 
the outstanding problems in particle physics today, since CP violation in the 
leptonic sector can be linked to leptogenesis and the matter-antimatter asymmetry of the 
Universe \cite{Endoh:2002wm}. 
Measuring the value of $\dcp$ can also provide valuable insights 
into new physics beyond the Standard Model, since CP violation can arise 
in various models of neutrino mass generation through complex couplings of 
neutrinos to other particles, or through
complex vacuum expectation values \cite{Branco:2011zb}. 

The measurement of $\dcp$ through oscillations of atmospheric/artificially-produced 
neutrinos is difficult using existing technology, and therefore new strategies have 
to be devised. Many interesting proposals exist in the literature for 
getting an evidence of CP violation and/or measuring $\dcp$ (for a non-exhaustive list, see 
Refs.~\cite{Winter:2006ce,Meloni:2012nk,cpmeasure}). 
In this paper, for the first time, we analyze actual UHE neutrino data to 
measure $\dcp$ and determine the source of astrophysical neutrinos.
In Ref.~\cite{Winter:2006ce}, the author discussed in detail the complementary 
nature of astrophysical 
and terrestrial neutrino experiments in CP studies. In that study (and more recently 
in Ref.~\cite{Meloni:2012nk}), data in the form of 
flavour ratios of observed neutrinos was used. In this work, we have analysed data from 
IceCube using a similar approach to get a hint about the value of $\dcp$. 

{\bf \underline{Astrophysical sources:}}
The data recorded by the IceCube telescope is the first evidence of extra-terrestrial events 
in the UHE range. These neutrinos can have their origin in extragalactic astrophysical 
sources like low power Gamma-Ray Burst (GRB) jets in stars \cite{Murase:2013ffa} or 
Active Galactic Nuclei (AGN) cores \cite{Stecker:2013fxa}.
{(Note however, that based on the data collected by the photon detectors 
Fermi, MAGIC, HESS, etc. in the 100 GeV -- TeV range, one can predict galactic sources of 
TeV neutrinos \cite{Ahlers:2013xia}.)} The energy of the 28 detected neutrino events are in the range $25 - 2000$ TeV. 
By tracing the hadronic origin \cite{Razzaque:2002kb} of these events, 
one can estimate the proton energies at their 
sources to be within $0.5-40$ PeV. Supernova Remnants (SNRs), AGNs, 
GRBs and other astrophysical sources can accelerate protons 
to such energies (and above) by the Fermi acceleration mechanism. 
The interactions of these protons with soft photons or matter from 
the source
can give UHE neutrinos through the following process: $p\gamma,pp
\rightarrow{\pi^{\pm}}X,\,~{\pi^{\pm}} \, \rightarrow \,{\mu^{\pm}}\nu_{\mu}(\bar\nu_{\mu}), 
\,~{\mu^{\pm}} \, \rightarrow \, {e^{\pm}}\bar\nu_{\mu}(\nu_{\mu})\nu_e(\bar\nu_e)$ \cite{piSsource} 
with a flux ratio of $\phi_{\nu_{e}}:\phi_{\nu_{\mu}}:\phi_{\nu_{\tau}} =
1:2:0$ (known as \piS\ process). Some of the muons, due to their light mass, can get cooled in the 
magnetic field quickly resulting in a 
neutrino flux ratio of $0:1:0$ (\muDS\ process). K-mesons, produced from $p\gamma$ interactions with a 
cross-section two orders of magnitude less than pions, will cool in the magnetic field of the source at 
higher energies compared to the pions. $K^{+}\rightarrow{\mu^{+}\bar{\nu^{\mu}}}\,$ is 
the dominant channel of neutrino production from cooled pions, with a branching fraction of 63\%, 
and with the same flux ratio as the pion decay \cite{Hummer:2011ms}.  
The $p\gamma$ interaction also produces high energy neutrons which would decay as 
$n\,\rightarrow p +e^-+\overline{\nu}_e$ to anti-neutrinos \cite{Moharana:2010su} with the flux 
ratio of $1:0:0$ (\nS\ process). The relative contribution of each channel depends on different parameters of the 
astrophysical source like the magnetic field, the strength of 
the shock wave and density of photon background \cite{Moharana:2011hh}. 
Apart from neutrinos these processes also produce high energy photons inside the source. 
Correlation of high energy photons with the UHE neutrinos can be considered as a signature 
of hadronic production inside the source. For example, a TeV neutrino can have an accompanying TeV photon at the source. However due to attenuation in the background radiation during propagation, PeV photons will have typical mean free path $\sim 10$ kpc \cite{Protheroe:1996si}. Thus, the associated photons of TeV neutrinos from extragalactic sources cannot reach earth. 

{\bf \underline{Analysis:}}
The main sources of astrophysical neutrinos in the energy range 10 TeV to 1200 TeV 
are the \piS, \muDS\ and \nS\ channels. 
However, the exact fraction of events in the detector from each of these sources 
is not known. Therefore, we have introduced relative fractions $k_1$, $k_2$ and $k_3$ 
for these three sources respectively, which are treated as free parameters in the problem 
subject to the normalization constraint $\sum k_i = 1$. In this study, we have 
not considered any other sub-dominant mode of neutrino production. 

Neutrinos oscillate during propagation, and our aim is to 
observe these oscillations. Given that the value of $E/L$ for such neutrinos 
is very large compared to the mass-squared differences between 
the neutrino mass states, we can only observe the average oscillation probability. 
Therefore, the probabilities take the simple form:
\beq
P(\nu_\alpha \to \nu_\beta) \equiv P_{\alpha\beta} = \sum_i |U_{\alpha_i}|^2 |U_{\beta i}|^2 ~.
\eeq
It is worth emphasizing that this oscillation probability depends only on 
the mixing angles and CP phase, but not on the mass-squared differences. 
Therefore, unlike in beam-based experiments where knowledge of the mass hierarchy 
is essential for CP sensitivity \cite{Prakash:2012az}, in this case we can 
(at least in principle) detect CP violation without suffering from the hierarchy 
degeneracy . 
Also 
note that $P_{\alpha\beta} = P_{\beta\alpha}$, therefore the probability 
can only be an even function of $\dcp$. As a consequence, we can treat neutrino 
and antineutrino oscillations on an equal footing. Another consequence of this is 
that every value of $\dcp$ allowed by the data will be accompanied by a degenerate 
solution $-\dcp$.

The distinction between tracks (which we 
assume to be $\nu_\mu$ CC events) and showers (which we assume to be 
$\nu_e$ or $\nu_\tau$, or $\nu_\mu$ NC events) is quite clear in the IceCube detector. 
We have folded the relative initial fluxes with the oscillation probabilities to 
get the relative number of events at the detector. Separation of muon events 
into CC and NC has been done using the ratio of the cross-sections at the 
relevant energy \cite{Gandhi:1995tf}.
We have done a simple analysis 
using the total events, instead of binning the data in energy and angle. Since the 
probability is 
almost independent of energy, this simplification is not expected to affect the 
analysis. This also allows us to neglect the effect of energy resolution. 
In Ref.~\cite{Aartsen:2013jdh}, the number of background events 
in the IceCube data set is estimated to be $10.6^{+5.0}_{-3.6}$. 
Of these, $6.0 \pm 3.4$ are expected to be veto penetrating atmospheric muons 
and $4.6^{+3.7}_{-1.2}$ are from the atmospheric neutrino background above energy 10 TeV.
The background assumed by IceCube could be an overestimation \cite{mena}, 
since (a) it has been estimated by extrapolating data, and (b) for atmospheric 
neutrinos the background has been calculated from 10 TeV while the events have been detected 
with lowest energy nearly 28 TeV. Therefore, we have used an estimate of 3 
background atmospheric muon tracks and 3.4 (the lower limit) background 
atmospheric neutrinos. IceCube have predicted a total of $8.4\pm4.2$ muon events and 
$6.6_{-1.6}^{+5.9}$ atmospheric neutrinos \cite{Aartsen:2014gkd} including the next set of neutrino events detected for the period of 988 days. Using 
the same analysis method we have taken the lowest limit of the backgrounds for our calculation. We have separated the background atmospheric neutrinos 
into tracks and showers using the same cross-sections as mentioned earlier.
These background events are subtracted from 
the data set in our analysis.
The neutrino background flux ratio has been included as 0:1:0, 
(which is close to 0.05:1:0 estimated in Ref.~\cite{deyoung2009neutrino})
since TeV range muons will penetrate all the way through the atmosphere.

In Refs.~\cite{Serpico:2005sz,Winter:2006ce,Meloni:2012nk}, the authors have 
proposed the use of the variable $R = {N_\mu}/{(N_e+N_\tau)} $
for the study of CP violation with astrophysical sources, where 
$N_\alpha$ is the flux of $\nu_\alpha+\bar{\nu}_{\alpha}$ at the detector. 
This variable helps by 
eliminating the overall source and detector-dependent normalization. Moreover, as 
studies of the up/down ratio as well as data/MC ratio in atmospheric 
neutrinos have shown, taking 
ratios of event rates can also reduce the effect of systematics \cite{ratios}. 
For our study, we have constructed a similar quantity $ \rho = {N_{track}}/{N_{shower}}$, 
with the flavour compositions of the track and shower events as mentioned before.

We have constructed the quantity $\rho^{data}$ using the IceCube data, and calculated 
$\rho^{theory}$ for a certain value of $\dcp$ as described above. Background events 
are subtracted from the data, as mentioned above.
The statistical $\chi^2$ is then computed using the Gaussian definition
\beq
\chi^2(\dcp) = \left( \frac{ \rho^{data}-\rho^{theory}}{\sigma_\rho} \right)^2 ~,
\eeq
where $\sigma_\rho = \sqrt{\rho^{data} (1-\rho^{data})/N}$~,
N being the number of data points \cite{lyonsstats}.
We have incorporated systematic effects using the method of pulls, with a systematic error 
of 5\%. Note that, we have marginalized the $\Delta \chi^2$ over the mixing angles 
($\theta_{23}$, $\theta_{13}$, $\theta_{12}$) within the ranges 
$\theta_{23}$ = 35$^\circ$ to 55$^\circ$, $\sin^{2} 2\theta_{13}$= 0.085 to 0.115 and 
$\theta_{12}$= 30$^\circ$ to 36$^\circ$ respectively. 
The priors added are $\sigma(\sin^{2} 2\theta_{13})$ = 0.01, 
$\sigma(\sin^2 2\theta_{23})$ = 0.1 and $\sigma(\sin^2 \theta_{12})$ = 0.0155.


{\bf \underline{Results:}}

\begin{figure}[t!]
\includegraphics[height=11cm,width=9.5cm]{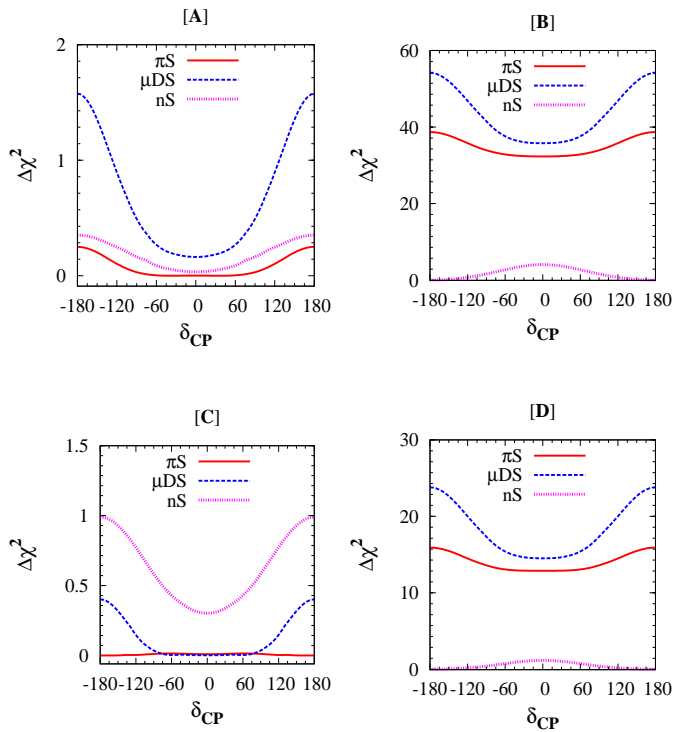}
\caption{\footnotesize Fit to $\dcp$ considering a single source i.e., 
\piS\ ($k_1=1, k_2=0, k_3=0$, solid red), 
\muDS\ ($k_1=0, k_2=1, k_3=0$, dashed blue), \nS\ ($k_1=0, k_2=0, k_3=1$, 
dotted magenta), considering all the events being from astrophysical environment. 
Panel [A]: Three-year data, without background; 
Panel [B]: Three-year data, with background; 
Panel [C]: Two-year data, without background; 
Panel [D]: Two-year data, with background. }
\label{sing_wbg-source}
\end{figure}

\begin{figure}
\includegraphics[height=5.5cm,width=9.cm]{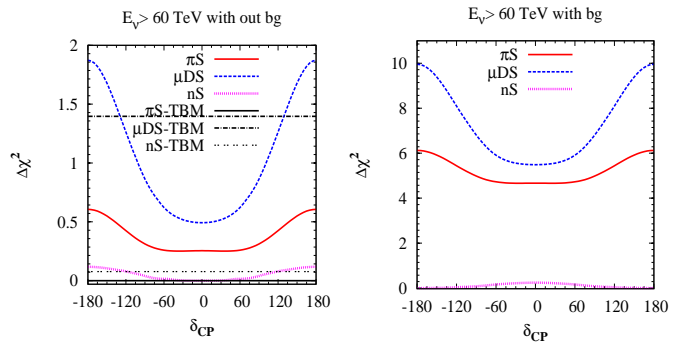}
\caption{Analysis for neutrinos within energy 60 TeV to 3 PeV.}
\label{with60}
\end{figure}

To demonstrate the impact of the origin of these astrophysical 
neutrinos on the precision of $\dcp$, we start with various possibilites, 
like, single , double or a combination of three sources as the origin.
First we show the fit to the data as a function of $\dcp$ for the 
single source assumption, in Fig. \ref{sing_wbg-source}. The upper row shows the results of our analysis of the full 
three-year data set. We have also included the results from analysing data from 
only the first two years (lower row) to show the improvement in results from 
additional data.

In the left panels we assume 
that all the events seen at IceCube are purely of astrophysical origin 
whereas in the right panels we include the effect of 
backgrounds. The latter is the realistic assumption. 
 From these figures 
we can see that, in case of no background the \piS\
source is 
favoured by the data as compared to the \nS\ and \muDS\ source(though the sensitivity 
is quite small, as $\Delta \chi^2$ is always $\textless$ 1.5). However, when we 
include the background, the scenario changes completely.
Pure \piS\ and pure \muDS\ sources are ruled out by the data 
at $> 3\sigma$, while the pure \nS\ source is favoured by data, though 
it is not sufficient to put any significant constraint 
on the value of $\dcp$. This has also been pointed 
out recently in Ref.~\cite{mena}.
This result can be  understood  qualitatively in the following way. 
In the 2nd column of Table~\ref{tab:werner} we have listed the 
theoretically calculated values  
for track by shower ratio for all the three sources keeping 
the oscillation parameters fixed at their tri-bimaximal (TBM) values 
($\theta_{23}=45^o$, $\theta_{13}=0^o$, $\sin^2\theta_{12}=
\frac{1}{3}$)\footnote{Due to the present non zero value of 
$\theta_{13}$, there will be deviations from the TBM values but 
as shown in Ref.~\cite{Meloni:2012nk}, this deviation is quite small.}
whereas the third column contains the experimental values of the track 
by shower ratio without and with backgrounds. 
 We can clearly see that 
for a pure signal, the track to shower ratio
for \piS\ is closest to the data. But the difference 
becomes quite high when backgrounds 
are taken under consideration, resulting in a very high $\Delta \chi^2$. A comparison of the upper and lower panels shows a marked 
increase in $\Delta \chi^2$. This shows the importance of additional data 
in both, excluding certain combinations of sources as well as constraining 
the value of $\dcp$.

 We have also done an analysis of the events in the energy range 60 TeV $< E <$ 3 PeV considering the 3 years of IceCube data. 
This is motivated by the fact that, this energy interval contains
the atmospheric muon background less than one. In this energy range there are 4 track events and 16 shower events with an atmospheric muon background of 0.435 and atmospheric
neutrino background of 2.365 \cite{Aartsen:2014gkd}. The result is plotted in Fig. \ref{with60}. In the left panel there is no background and in the right panel 
background has been considered. From the right panel we can see that we are still getting \nS\ as the favoured source whereas \piS\ and \muDS\ sources are excluded at more than
$2 \sigma$. This is due to tha fact that though the atmospheric muon background is less than one in this energy range but due to the presence of atmospheric neutrino
background \nS\ is getting preferred over \piS\ source.  This can bee seen from the left panel where no background is considered. There we can note that
the data agrees with the final flavor ratio 1:1:1 i.e it favours the \piS\
source over \nS\ source marginally
when TBM mixing is assumed. But when we vary the oscillation parameters in their allowed $3 \sigma$ range then due to the deviation from TBM, \nS\ is getting slightly
preferred over \piS\ .

\begin{figure}[t!]
\includegraphics[height=6cm,width=7.cm]{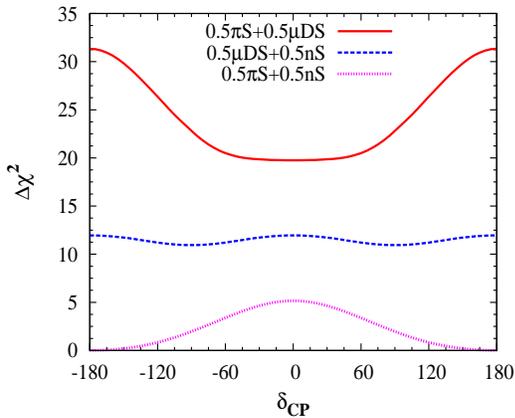}
\caption{\footnotesize Fit to $\dcp$ considering contribution from two 
sources at a time, in equal proportion i.e., 
$k_1=k_2=0.5, k_3=0$ (solid, red), $k_1=0, k_2= k_3=0.5$ (dashed, blue), 
$k_1=k_3=0.5, k_2=0$ 
(dotted, magenta) for both old {\bf {a}} and new {\bf B}.}
\label{doub-source}
\end{figure}
\begin{figure}[t!]
\includegraphics[height=6cm,width=8.cm]{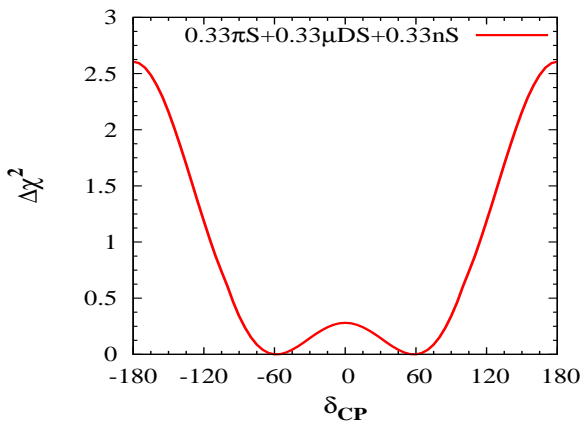}
\caption{\footnotesize Fit to $\dcp$ considering equal contribution from all the 
sources, i.e.,  $k_1= k_2= k_3=0.33$ for both old {\bf {a}} and new {\bf B}.}
\label{equal-source}
\end{figure}

In Fig.~\ref{doub-source} and Fig.~\ref{equal-source} we show the fit to the data 
when neutrinos are coming from two/all the three sources 
respectively, with equal contributions. These results are for the full data set, and 
backgrounds have been included in generating these plots. 
In Fig.~\ref{doub-source}, we find only the combination of \piS\ and \nS\ 
neutrinos are allowed at $3\sigma$ level. We also see that the CP dependence is 
maximum if the neutrinos come from the combination of \piS\ and 
\muDS\ modes. The data may also rule out one-third of $\dcp$ values 
(approximately $-60^\circ$ to $60^\circ$) at $\sim 2\sigma$. The poor sensitivity 
from \nS\ neutrinos is the reason why the combination 
of \piS+\muDS\ in Fig.~\ref{doub-source} 
has a higher $\chi^2$ than the combinations involving \nS.
When we 
consider equal contributions from all these channels (Fig.~\ref{equal-source}), 
we find that the data favours the first and fourth quadrants of $\dcp$ at $1\sigma$.


\begin{table}
\renewcommand{\arraystretch}{1.5}
 \begin{tabular}{|c|c||c|}
  \hline
  Source & $\frac{N_{track}}{N_{shower}}$(Calculated) & $\frac{N_{track}}{N_{shower}}$(Data) \\
  \hline
  \piS & 0.30 &  \\
       &      &  8/28=0.287(Without background) \\
  \muDS & 0.38 &  \\
        &      & 0.06(With background)   \\
  \nS & 0.18 & \\
  \hline
 \end{tabular}
\renewcommand{\arraystretch}{1}
\caption{\footnotesize Theoretical values of track by shower ratio 
for all the three sources along with experimental values with and without background.}
\label{tab:werner}
\end{table}

\begin{figure}[t!]
\includegraphics[height=6.cm,width=10.cm]{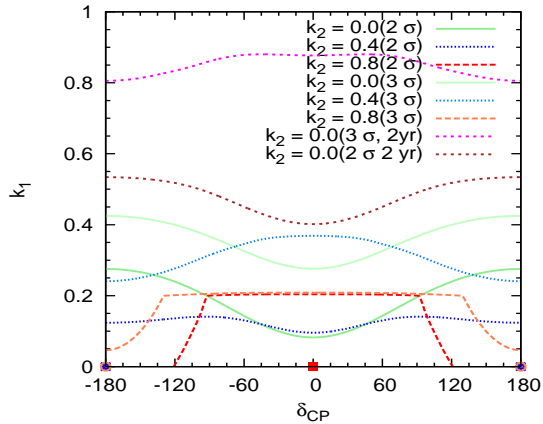}
\caption{\footnotesize Contour plots for allowed region in the $k_1-\dcp$ plane, 
for three representative values of $k_2$. The points marked in the respective colours 
indicate the best-fit point with new IceCube data.}
\label{cont_k1}
\end{figure}

\begin{figure}[t!]
\includegraphics[height=6.cm,width=10.cm]{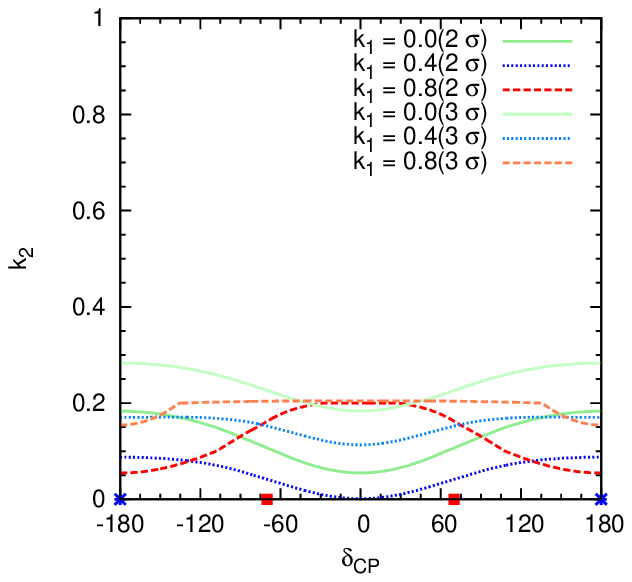}
\caption{\footnotesize Contour plots for allowed region in the $k_2-\dcp$ plane, 
for three representative values of $k_1$. The points marked in the respective colours 
indicate the best-fit point.}
\label{cont_k2}
\end{figure}

\begin{figure}[t!]
\includegraphics[height=6.cm,width=10.cm]{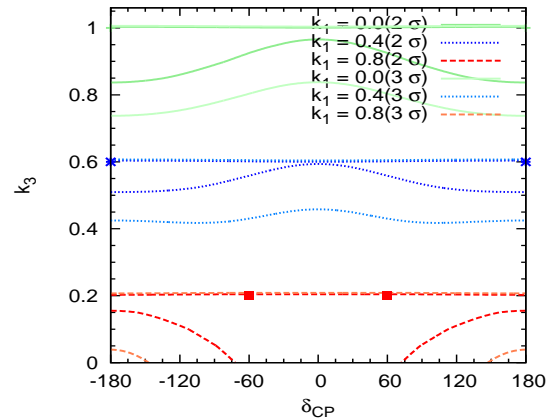}
\caption{\footnotesize Contour plots for allowed region in the $k_3-\dcp$ plane, 
for three representative values of $k_1$. The points marked in the respective colours 
indicate the best-fit point.}
\label{cont_k3}
\end{figure}

We have then performed a check to constrain the astrophysical 
parameters $k_i$ vs $\dcp$ using the IceCube 
data, by plotting the allowed countours in the $k_i-\dcp$ 
plane. In Fig.~\ref{cont_k1}, 
we have showed the $2\sigma$ (light)
and $3\sigma$ (dark) contours in the $k_1-\dcp$ plane for 
three fixed values of $k_2$. The best-fit point indicated by the data has been 
marked with a red dot. 
We see that the data favours a smaller value of $k_1$ and 
larger values of $k_2$ and $k_3$. 
Similarly, Fig.~\ref{cont_k2} shows that for a given value of $k_1$, the 
data disfavours the \muDS\ process (small value of $k_2$) but favours the \nS\ process 
(large value of $k_3$). Likewise, Fig.~\ref{cont_k3} shows the data favouring the
largest possible value of $k_3$ allowed by the normalization condition. These features 
can be understood from Fig.~\ref{sing_wbg-source}, where we see that the data prefers the 
\nS\ source. From these contours, we may draw certain contraints on the 
astrophysical sources most favoured. In particular if we obtain a good prior 
on $\dcp$ from other experiments, then the most favoured ratio of $k_1$, $k_2$ and 
$k_3$ may be obtained. Alternately, if we obtain a better picture of the 
sources of the IceCube events, a refined and constrained range on  $\dcp$ would 
be predicted.


 To show the statistical improvement of the 3 year data over 2 year data, 
in Fig.~\ref{cont_k1} we have also plotted the $2\sigma$ and $3\sigma$ contours 
for the latter for $k_2=0$. Here we can clearly see that for $\delta_{CP}=0$, 
3 year data can exclude $73\%$($91.5\%$) of $k_1$ values at $3\sigma$($2\sigma$) 
where as the 2 year data can only rule 
out $13\%$($61\%$) of $K_1$ values at $3\sigma$($2\sigma$). For $\delta_{cp}=\pi$ 
the exclusion percentages are $58\%$($73\%$) at $3\sigma$($2\sigma$) for 3 years
 and $20\%$($48\%$) at $3\sigma$($2\sigma$) for 2 years. One can undestand this 
 qualitatively from the \piS\ curve of Fig.~\ref{sing_wbg-source} showing a significant 
 improvement in the 
$\Delta\chi^2$ with 3 years of data compared to 2 years.


{\bf \underline{Conclusion:}}
In this work, we have analyzed the first IceCube data on TeV-PeV scale neutrinos. 
We have used the flux ratios 
of the three neutrino flavours to put constraints on $\dcp$. We find that the results 
depend strongly on the source of the neutrinos. 
After taking into account the effect of backgrounds, we find that the \nS\ source 
of neutrinos is favoured by the data. Depending on the particular combination of sources for these neutrinos, 
current data can only hint at the allowed region of the $\dcp$ range. However, 
we have shown that additional data gives a remarkable improvement in results, which 
underlines the importance of future data from IceCube.
We have also put constraints on the astrophysical parameters $k_1$, $k_2$ and $k_3$ 
that determine which of the 
modes of neutrino production is more close to the data. In fact if $\dcp$ is measured by other 
experiments, then IceCube
data can be used to determine the production mechanism of these neutrinos. 
Similar analyses can also be carried 
out for other parameters related to neutrino physics and astrophysics. 

{\bf \underline{Acknowledgements:}}
We thank the Workshop on High Energy Physics 
Phenomenology 2013 (WHEPP13), where this work was 
done. We also thank Amol Dighe, Raj Gandhi and Srubabati 
Goswami 
for useful comments and suggestions. RM likes to thank University of Johannesburg for its hospitality, where some of this paper work was done.


\end{document}